\documentclass[a4paper]{jpconf}
\usepackage{graphicx}
\begin{document}
\title{Quantum Phase Transition in a Graphene Model}

\author{Simon Hands$^1$, Costas Strouthos$^{2}$}

\address{$^1$ Department of Physics, Swansea University, Singleton Park, Swansea SA2 8PP, UK}
\address{$^2$ Department of Mechanical Engineering, 
University of Cyprus, Nicosia 1678, Cyprus} 

\ead{s.hands@swansea.ac.uk, strouthos@ucy.ac.cy}

\begin{abstract}
We present results for the equation of state of a graphene-like model in an effort to understand the 
properties of its quantum phase transition. The $N_f$ fermion species interact through a three dimensional
instantaneous Coulomb potential. Since there are no reliable analytical tools that work for all values of
$N_f$ and the coupling constant $g$ , we rely on Monte Carlo simulations to calculate the critical 
properties of
the model near the phase transition. We consider the four-component formulation for the fermion ﬁelds,
which arises naturally as the continuum limit of the staggered fermion construction in (2+1) dimensions.
In the limit of infinitely strong Coulomb interaction, the system undergoes a quantum phase transition
at a critical number of fermion species $N_{fc} \approx 4.7$. 
We also calculate the values of the critical exponents at the quantum phase transition.
\end{abstract}

\section{Introduction}
There has been considerable recent interest in graphene sparked by its
discovery and subsequent experimental study. In brief, for a carbon monolayer having one mobile
electron per atom,
a simple tight-binding model shows that the spectrum of low-energy excitations
exhibits a linear dispersion relation centred on zeroes located at the six
corners of the first Brillouin
zone (see e.g. \cite{Jackiw:2007rr}). Using a linear transformation among
the fields at two independent zeroes it is possible to recast the Hamiltonian
in Dirac form with $N_f=2$ flavors of
four-component spinor $\psi$, the counting of the massless
degrees of freedom coming from 2
carbon atoms per unit cell $\times$ 2 zeroes per zone $\times$
2 physical spin components per electron. Electron propagation in the graphene
layer is thus relativistic, albeit at a speed $v_F\approx c/100$. Although this is a factor of 
$100$ slower than the speed of light in vacuum, it is still much faster than the speed of 
electrons in an ordinary conductor.
The implications for the high mobility of the resulting charge carriers (which may
be negatively-charged ``particles'' or positively-charged ``holes'' depending on
doping) is the source of the current excitement. The stability of the
zero-energy points is topological in origin, as emphasised by
Creutz~\cite{Creutz:2007af}.

While the above considerations apply quite generally, a realistic model of
graphene must
incorporate interactions between charge carriers. One such model, due to Son
\cite{Son:2007ja}, has $N_f$ massless fermion flavors propagating in the
plane, but interacting via an
instantaneous 3$d$ Coulomb interaction. In Euclidean metric and static gauge
$\partial_0A_0=0$ the action reads
\begin{equation}
S_1=\sum_{a=1}^{N_f}\int dx_0d^2x(\bar\psi_a\gamma_0\partial_0\psi_a
+v_F\bar\psi_a\vec\gamma.\vec\nabla\psi_a+iV\bar\psi_a\gamma_0\psi_a)
+{1\over{2e^2}}\int dx_0d^3x(\partial_i V)^2,
\label{eq:model}
\end{equation}
where $e$ is the electron charge, $V\equiv A_0$, and the $4\times4$ Dirac matrices satisfy
$\{\gamma_\mu,\gamma_\nu\}=2\delta_{\mu\nu}$, $\mu=0,1,2,3$.
In our notation
$\vec x$ is a vector in the 2$d$ plane while the index $i$ runs over all three
spatial directions. 
In the large-$N_f$ limit the dominant quantum correction
$\Pi(p)$
comes from a vacuum polarisation fermion- antifermion loop and the resummed
$V$ propagator becomes
\begin{equation}
D_1(p)=(D_0^{-1}(p)-\Pi(p))^{-1}
=\left({{2\vert\vec p\vert}\over e^2}+{N_f\over8}{{\vert\vec
p\vert^2}\over{(p^2)^{1\over2}}}\right)^{-1},
\label{eq:D_gr}
\end{equation}
where $p^2=(p_0,\vec p)^2\equiv p_0^2+v_F^2\vert\vec p\vert^2$, and $D_0(p)$ corresponds
to the classical propagation of V.
In either the strong
coupling or large-$N_f$ limits $D_1(p)$ is thus dominated by 
$\Pi(p)$, the relative importance of the original Coulomb interaction being
governed by a parameter $\lambda\equiv\vert\Pi/D_0\vert_{p_0=0}$. 

The chiral symmetry breaking, due to the spontaneous condensation of particle - hole
pairs, is signalled by an order parameter $\langle\bar\psi\psi\rangle\not=0$. 
Physically the most important outcome is the generation of a gap in the fermion
spectrum, implying the model is an insulator. Son \cite{Son:2007ja} postulates that this
insulating phase exists in the corner of the phase diagram corresponding to
large $e^2$ and small $N_f$, and in particular that the insulator-conductor
phase transition taking place at $N_f=N_{fc}$ in the strong-coupling limit
$e^2\to\infty$ is a novel quantum critical point. The value of $N_{fc}$, and the
issue of whether it is greater or less than the physical value $N_f=2$,
must be settled by a non-perturbative calculation. A recent estimate obtained by
a renormalisation group treatment of radiatively-induced four-fermion contact
interactions, is $N_{fc}=2.03$ \cite{Drut:2007zx}.

The proposed physics is very reminiscent of the three dimensional Thirring model, which 
is analytically tractable at large $N_f$, but exhibits spontaneous chiral symmetry breaking 
at small $N_f$ and strong coupling \cite{DelDebbio:1997dv,DelDebbio:1999xg}. 
Arguably the Thirring model is the simplest field theory of fermions requiring a
computational solution:
the location of the phase transition at $N_f=N_{fc}$ in the strong coupling
limit has recently been determined by lattice simulations to be
$N_{fc}=6.6(1)$~\cite{Christofi:2007ye}. The apparent similarity of the two
systems has led us to propose a Thirring-like model pertinent to graphene, with
Lagrangian 
\begin{equation}
S_2= \sum_{a=1}^{N_f}\int dx_0d^2x  \left[ \bar\psi_a\gamma_\mu\partial_\mu\psi_a
+iV\bar\psi_a\gamma_0\psi_a+{1\over{2g^2}}V^2 \right].
\label{eq:model2}
\end{equation}
As for (\ref{eq:model}) we assume a large-$N_f$ limit to
estimate the dominant vacuum polarisation correction; the resultant propagator
for $V$ is
\begin{equation}
D_2(p)
=\left({1\over g^2}+{N_f\over8}{{\vert\vec
p\vert^2}\over{(p^2)^{1\over2}}}\right)^{-1}.
\label{eq:D_Th}
\end{equation}
In the strong-coupling or large-$N_f$ limits,
$D_2$ coincides with $D_1$,
implying that the fermion interactions are equivalent.
It is also the case that $\lim_{p\to\infty}
D_2(p)=\lim_{\lambda\to\infty}D_1(p)$.
This last limit is
important because critical behaviour in the Thirring  model is governed by a
UV-stable fixed point of the renormalisation group~\cite{DelDebbio:1997dv}.
We thus expect predictions made with the model (\ref{eq:model2}), and in
particular critical behaviour such as the
value of $N_{fc}$,
to be generally valid for Son's model (\ref{eq:model}) in the limit of large
$\lambda$.

\section{Lattice Simulations}
The lattice model studied in this paper is related to the lattice Thirring
model studied in \cite{DelDebbio:1997dv}, except that in this case we used 
the compact formulation for the auxiliary boson field. The simulations were
performed with the standard hybrid molecular dynamics algorithm.
\begin{figure}[]
\begin{center}
\includegraphics[width=17pc]{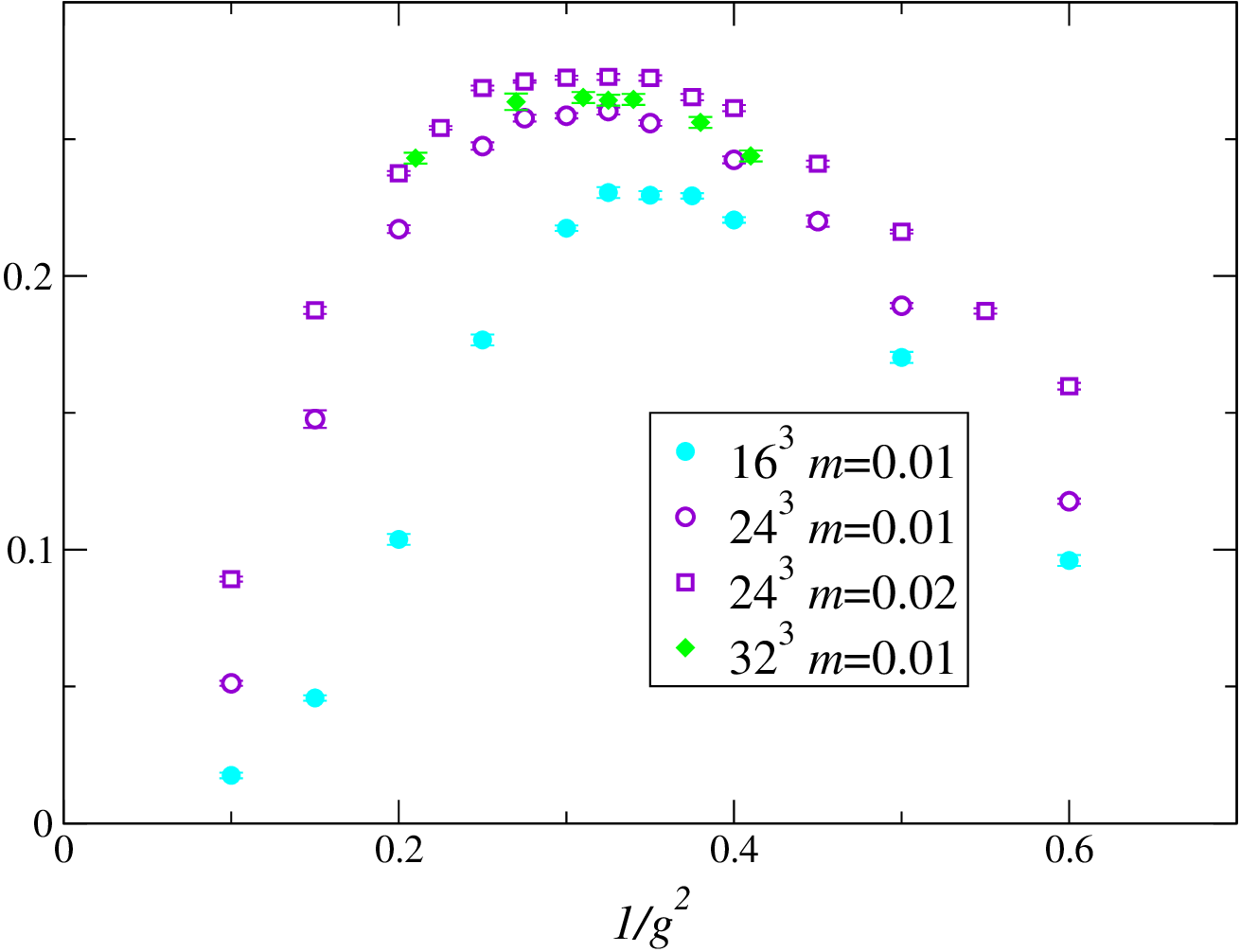}
\caption{\label{gr:beta_peak_Nf2} $\langle \bar{\psi}\psi \rangle$ vs. $1/g^2$ near $g_{\rm peak}$ for $N_f=2$.}
\end{center}
\end{figure}
The novelty of \cite{Christofi:2007ye} is that it is the first study of the Thirring model
by lattice simulations
in the strong-coupling limit $g^2\to\infty$. Since we aim to repeat the strategy
here we should discuss how this was done. First, note that the vacuum
polarisation calculation leading to the results (\ref{eq:D_gr},\ref{eq:D_Th})
does not go through in quite the same way for the lattice regularised model; 
rather, there is an additive correction which is
momentum independent and UV-divergent:
\begin{equation}
\Pi^{\rm latt}(p)=\Pi^{\rm cont}(p)+g^2J(m),
\end{equation}
where $J(m)$ comes from incomplete cancellation of a lattice tadpole diagram
\cite{DelDebbio:1997dv}. This extra divergence not
present in the continuum treatments can be absorbed by a wavefunction
renormalisation of $V$ and a coupling constant renormalisation
\begin{equation}
g_R^2={g^2\over{1-g^2J(m)}}.
\end{equation}
In the large-$N_f$ limit
we thus expect to find the strong coupling limit of the lattice model at
$g_R^2\to\infty$ implying $g^2\to g^2_{\rm lim}$.
For $g^2>g^2_{\rm
lim}$ $D_{\rm latt}(p)$ becomes negative, and $S_{\rm latt}$ no longer describes a
unitary theory.

\begin{figure}[h]
\begin{minipage}{17pc}
\includegraphics[width=17pc]{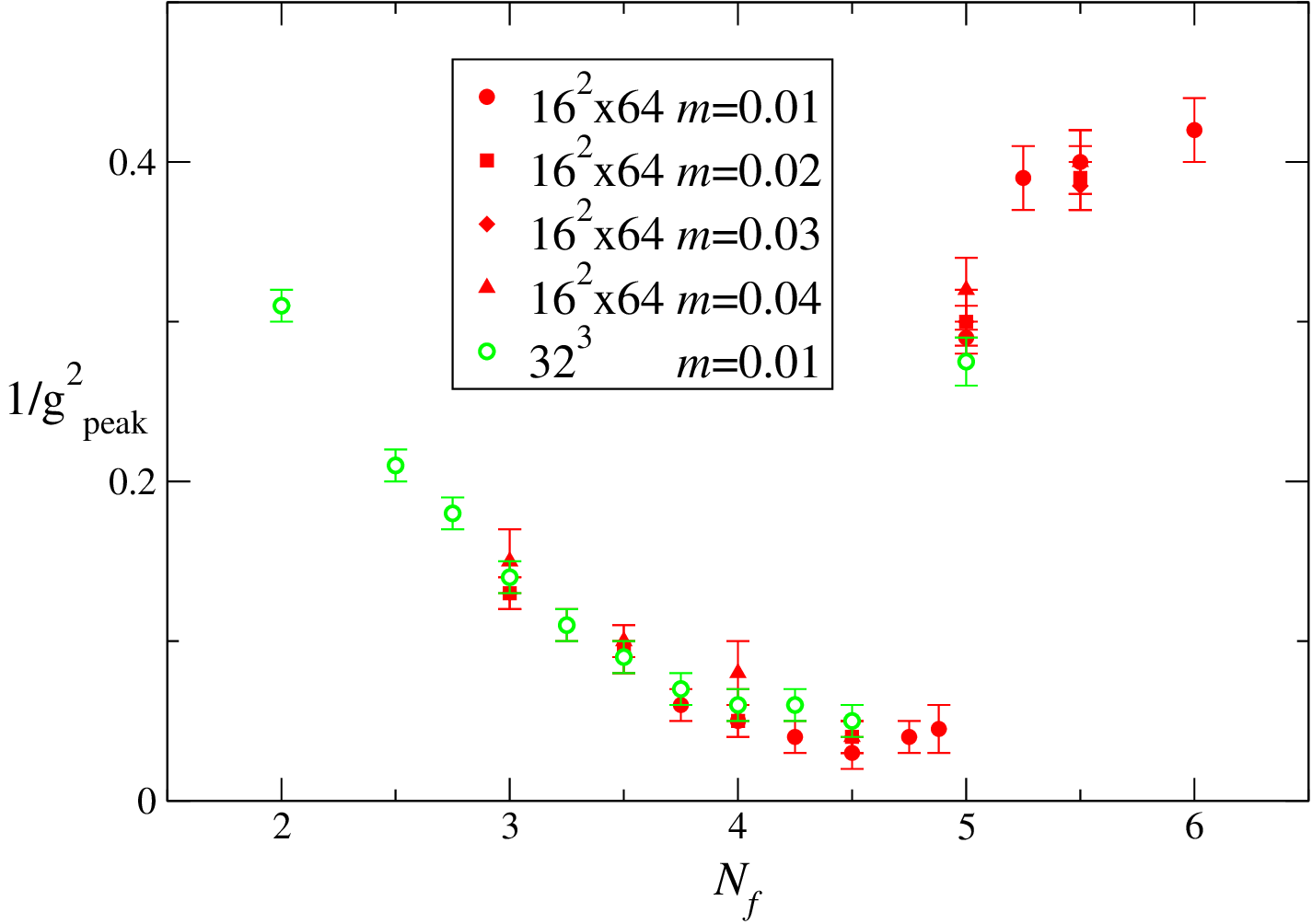}
\caption{\label{gr:beta_peak}$1/g^2_{\rm peak}$ vs. $N_f$.}
\end{minipage}\hspace{2pc}%
\begin{minipage}{17pc}
\includegraphics[width=17pc]{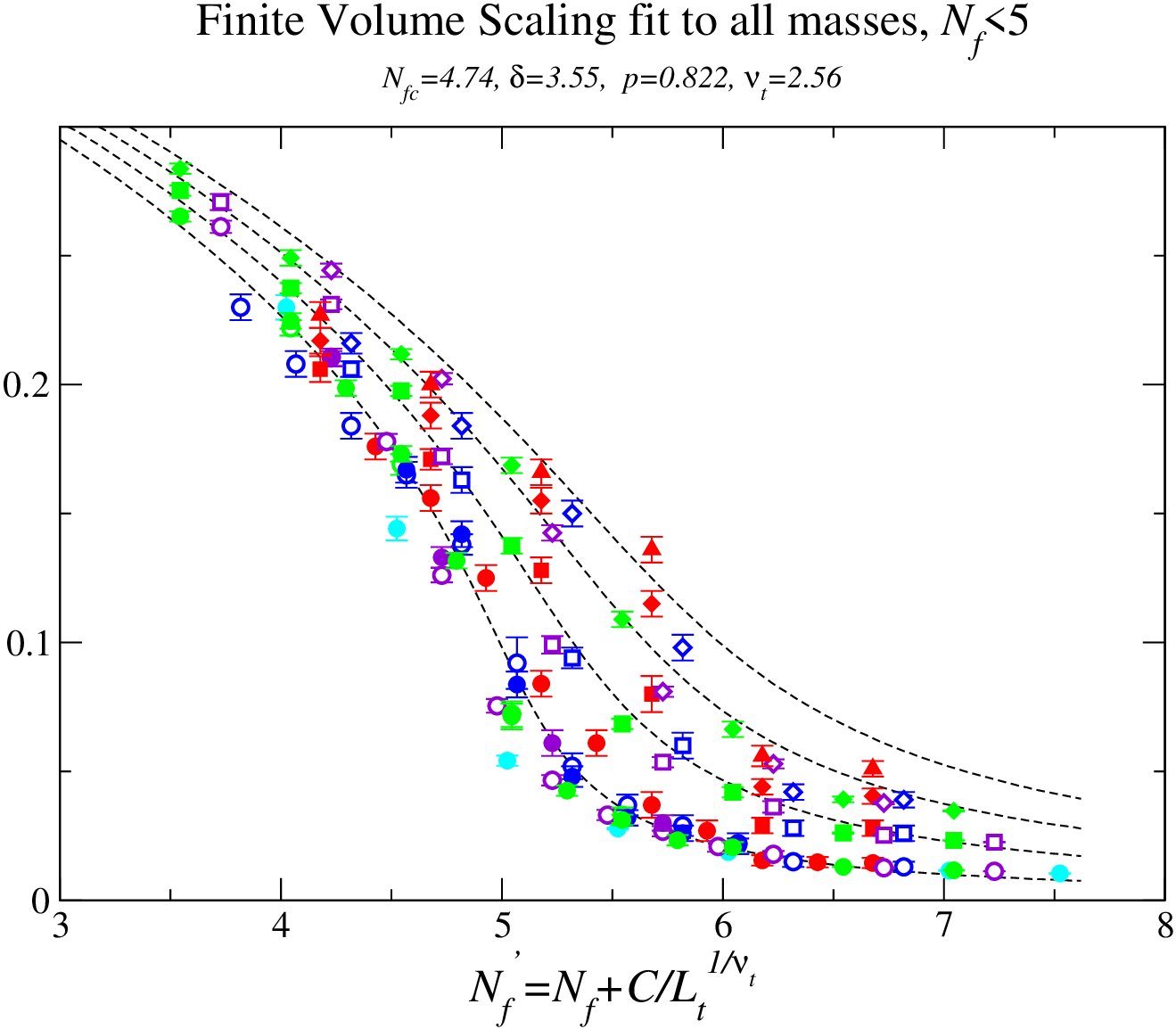}
\caption{\label{gr:eos}Equation of State fits to data generated with different fermion masses 
on different lattice sizes.}
\end{minipage}
\end{figure}

Away from the large-$N_f$ limit, where chiral symmetry may be spontaneously
broken, there is no analytical criterion for identifying
$g^2_{\rm lim}$; however in this case a numerical calculation of
$\langle\bar\psi\psi\rangle$ shows a clear peak at $g^2=g^2_{\rm peak}$,
whose location is approximately independent of both volume and $m$, indicating
an origin at the UV scale \cite{Christofi:2007ye}.
Fig.~\ref{gr:beta_peak_Nf2} exemplifies this behaviour in the
lattice model with $N_f=2$ on system
volumes $L^3$: for $L\geq24$ we identify $1/g^2_{\rm peak}\simeq0.3$. Since for
orthodox chiral symmetry breaking the magnitude of the condensate is expected to
increase monotonically with the coupling strength, we interpret the peak as the
point where unitarity violation sets in, ie.
$g^2_{\rm lim}\approx g^2_{\rm peak}$.
We use simulations performed at $g^2=g^2_{\rm lim}$
to explore the strong coupling limit, and find clear evidence for a chiral
symmetry restoring phase transition at a well-defined $N_{fc}$. As shown in Fig.~\ref{gr:beta_peak}
$1/g^2_{\rm peak}$ decreases as $N_f$ increases from $2$ to $4.75$. Near $N_f \approx 4.75$ the curve reaches 
a minimum, implying a significant change in the strong coupling behavior of the model. The rapid change 
in the value of the condensate at $N_f \approx 4.9$ implies a signifacant change in the strong coupling
behavior of the model. This can be a phase transition that separates a 
chirally broken from a chirally symmetric phase. 
Since our model has anisotropic interactions in the spatial and temporal directions the correlation
lengths near the transition diverge with different exponents,  $\nu_s$ in the spatial directions, and 
$\nu_t$ in the temporal direction. In this case a modified hyperscaling relation holds:
\begin{equation}
\nu_t + (d-1)\nu_s = \gamma + 2\beta.
\end{equation}
We carefully monitored the finite size effects in the spatial directions   
by comparing the values of the condensate at $m=0.01$, $g^2_{\rm peak}(N_f)$ extracted from simulations on  
$16^2 \times 48$ and $24^2 \times 48$ 
lattices. Our results show that finite $L_s$ effects are negligible for $L_s \geq 16$. 
We fitted the data generated on several lattices with $L_t=16$, $L_s=48, 64$ and $m=0.01,...,0.04$ 
to the following RG-inspired equation of state \cite{DelDebbio:1997dv}:
\begin{equation}
m=A[(N_f-N_{fc})+CL_t^{-{1\over\nu_t}}]\langle\bar\psi\psi\rangle^p
+B\langle\bar\psi\psi\rangle^\delta,
\end{equation}
where $p \equiv \delta - \beta^{-1}$. The results show clearly that the model undergoes 
a second order quantum phase transition 
and the values of the most significant parameters are: $N_{fc}=4.74$, 
$\delta=3.55$, $p=822$, and $\nu_t=2.56$. The data and the fitted curves are shown in Fig.~\ref{gr:eos}.
The value of $N_{fc}=4.74$ is close to the value of $N_f \approx 4.9$, where a rapid change in 
the $1/g^2_{\rm peak}$ vs $N_f$ occurs.  

\section{Summary}
We presented results from a non-perturbative Monte Carlo study of a Thirring-like model pertinent to graphene.
Our analysis of its equation of state at strong coupling reveals that the model undergoes a second 
order quantum phase transition at $N_{fc} \approx 4.74$, 
which is greater than the physical $N_f=2$ value. This result implies that freely suspended graphene 
may be an insulator for couplings $g$ larger than a critical coupling $g_c$.


\section*{References}
\begin{thebibliography}{9}


\bibitem{Jackiw:2007rr}
  R.~Jackiw and S.~Y.~Pi, Phys. Rev. Lett. {\bf 98} (2007) 266402.
  %``Chiral Gauge Theory for Graphene,''
  % arXiv:cond-mat/0701760.
  %%CITATION = COND-MAT/0701760;%%

\bibitem{Creutz:2007af}
  M.~Creutz,
  %``Four-dimensional graphene and chiral fermions,''
  JHEP {\bf 0804} (2008) 017.
  %[arXiv:0712.1201 [hep-lat]].
  %%CITATION = JHEPA,0804,017;%%

\bibitem{Son:2007ja}
  D.T.~Son,
  %``Quantum critical point in graphene approached in the limit of infinitely
  %strong Coulomb interaction,''
  Phys.\ Rev.\  B {\bf 75} (2007) 235423.
  %arXiv:cond-mat/0701501.
  %%CITATION = PHRVA,B75,235423;%%

\bibitem{Drut:2007zx}
  J.E.~Drut and D.T.~Son,
  %``Renormalization group flow of quartic perturbations in graphene: Strong
  %coupling and large-N limits,''
  Phys.\ Rev.\  B {\bf 77} (2008) 075115.
  %[arXiv:0710.1315]
  %%CITATION = PHRVA,B77,075115;%%

\bibitem{DelDebbio:1997dv}
  L.~Del Debbio, S.J.~Hands and J.C.~Mehegan,
  %``Three-dimensional Thirring model for small N(f),''
  Nucl.\ Phys.\  B {\bf 502} (1997) 269.
  %[arXiv:hep-lat/9701016];\\
  %%CITATION = NUPHA,B502,269;%%

\bibitem{DelDebbio:1999xg}
  L.~Del Debbio and S.J.~Hands,
  %``The three dimensional Thirring model for N(f) = 4 and N(f) = 6,''
  Nucl.\ Phys.\  B {\bf 552} (1999) 339; 
  %[arXiv:hep-lat/9902014];\\
  %%CITATION = NUPHA,B552,339;%%
%\bibitem{Hands:1999id}
  S.J.~Hands and B.~Lucini,
  %``The phase diagram of the three dimensional Thirring model,''
  Phys.\ Lett.\  B {\bf 461} (1999) 263.
  %[arXiv:hep-lat/9906008].
  %%CITATION = PHLTA,B461,263;%%

\bibitem{Christofi:2007ye}
  S.~Christofi, S.J.~Hands and C.~Strouthos,
  %``Critical flavor number in the three dimensional Thirring model,''
  Phys.\ Rev.\  D {\bf 75} (2007) 101701.
  %%CITATION = PHRVA,D75,101701;%%

%\bibitem{BB} C.J. Burden and A.N. Burkitt, Europhys. Lett. {\bf3} (1987) 545.

%\bibitem{HMD}
%S. Gottlieb, W. Liu, D. Toussaint, R.L. Renken and R.L. Sugar, Phys. Rev. {\bf
%D35} (1987) 2531.

%\bibitem{HMC}
%S. Duane, A.D. Kennedy, B.J. Pendleton and D. Roweth, blah blah.

%\bibitem{Barbour:1998yc}
%  I.M.~Barbour, N.~Psycharis, E.~Focht, W.~Franzki and J.~Jers\'ak,
  %``Strongly coupled lattice gauge theory with dynamical fermion mass
  %generation in three dimensions,''
%  Phys.\ Rev.\  D {\bf 58}, 074507 (1998)
  %%CITATION = PHRVA,D58,074507;%%



%\bibitem{pennington}
%M.R. Pennington and D. Walsh, Phys. Lett. {\bf B253} (1991) 246.


\end{thebibliography}
\end{document}